\documentstyle[12pt, a4,fleqn]{article}
\begin{document}
\topmargin= -20mm
\textheight= 230mm
\baselineskip = 0.33 in
 \begin{center}
\begin{large}
 {\bf { BOUND STATES 
 
 OF 
 
 1+1 DIMENSIONAL FIELD THEORIES }}
\end{large}

\vspace{2cm}

 T. Fujita, M. Hiramoto and H. Takahashi

Department of Physics, Faculty of Science and Technology  
  
Nihon University, Tokyo, Japan

\vspace{3cm}

{\large ABSTRACT} 

\end{center}

We discuss the bound states of the massive Thirring model. Here, 
the periodic boundary condition equations for the Bethe ansatz solutions 
are numerically solved. It is found that the massive Thirring model 
has only one bound state and the bound state spectrum as the function of 
the coupling constant agrees with that of infinite momentum frame 
prescription by Fujita and Ogura. Boson boson states (2p$-$2h states) 
appear only as the continuum spectrum without making any bound states. 

Further, the finite size correction to the vacuum energy is estimated. 
The evaluated central charge is found to be close to unity. 

\newpage
 
\begin{enumerate}
\item{\large Introduction}

The sine-Gordon field theory or the massive Thirring model is believed 
to be solved exactly. Dashen, Hasslacher and 
Neveu presented their solutions to the quantum 
 sine-Gordon model [1].  Although they use 
semiclassical approximations, they consider their solutions to be $exact$. 
This spectrum is translated into the massive Thirring model and 
the bound state mass $ {\cal M} $ (vector boson)  is written as 
$$ {\cal M} = 2m \sin {\pi\over 2} {n\over{(1+{2g_0\over{\pi}}) }} 
 \eqno{(1.1)} $$
where $n$ is an integer and runs from 1 to $(1+{2g_0\over{\pi}}) $. 
 $m$ is the fermion mass of the massive Thirring model. $g_0$ is the 
coupling constant with Schwinger's normalization. 

Further, this spectrum is confirmed by the Bethe ansatz solution [2]. 
This was 
very important since the Bethe ansatz wave function is indeed exact. 
In their paper, Bergknoff and Thacker presented their solutions 
of the massive Thirring model based on the $string$ hypothesis when 
they solve equations of the periodic boundary conditions (PBC) from 
Bethe ansatz wave functions. 

However, Fujita and Ogura [3] have recently presented their solutions 
of the massive Thirring model employing infinite momentum frame 
prescription. Their spectrum is quite different from eq. (1.1).  
There is only one bound state. However, the bound state energy is 
rather close to the lowest energy of eq.(1.1). The deviation is 
about  10 $\sim$ 20 \% from each other depending on the coupling constant. 
The boson mass  $ {\cal M} $  is given as 
$$ { \tan \alpha  \over{{\pi\over 2}-\alpha}} = 
{g\over{\pi}} \left[ 1+{1\over
{\cos^2 \alpha}}(1-{g\over 4\pi}) \right] \eqno{(1.2)} $$
where   
the boson mass  $\cal M$ is related to $\alpha$  as,
$$ {\cal M} = 2m \cos \alpha .  $$ 
$g$ is a coupling constant of the massive Thirring model 
with Johnson's normalization.  

It turns out that the solution eq.(1.2) 
 has all the proper behaviors of the weak and strong coupling limits. 
Instead, if one checks eq.(1.1) carefully, then one sees that 
the semiclassical result of eq.(1.1)  
 does not have a proper weak coupling limit. There, the important point 
 is that  one has to 
take into account  current regularizations in a correct way [3]. 

On the other hand, the Bethe ansatz wave function is 
well known to be exact.  
  This is a strong reason why people 
have believed for almost two decades 
that the bound state spectrum obtained from 
the semiclassical approximation is exact in spite of the fact that they took 
into account only the lowest quantum fluctuations in the path integral. 

In this report, we reexamine the Bethe ansatz solutions for the massive 
Thirring model and discuss problems  in the treatment by 
Bergknoff and Thacker [2]. 
In particular, we show that the $string$ configurations taken by Bergknoff 
and Thacker  do not satisfy the PBC equations.  
The reason why they have to introduce the $string$ picture is because they 
solve the PBC equations for the density of states. Therefore, they 
could not determine proper rapidities for the positive energy particles. 

Here, we have solved the PBC equations 
numerically. We consider  a few hundred  particles  to a few thousand 
 particles to make a vacuum. Then, we make one particle-one hole pairs, 
two particle two hole pairs and so on. It is found that there is only 
one bound state for one particle-one hole ($1p-1h$) 
configuration. There is no 
bound state for two particle two hole cases. Further, the bound state 
energy calculated from the Bethe ansatz PBC equations 
turns out to be consistent with that of Fujita-Ogura's solution [eq. (1.2)]  
though we can solve only a limited region of the coupling constant. 

Further, we find the boson boson scattering states in $2p-2h$ configurations.  
Here, it is important to note that the boson boson scattering states have 
rapidity variables which are all real.  
Therefore, there is no $string-$like 
solution which satisfies the PBC equations.  

\vspace{2cm}

\item{\large Solutions of PBC equations}

The massive Thirring model is a 1+1 dimensional field theory with 
current current interactions. Its lagrangian density can be written as 
$$  {\cal L} =  \bar \psi ( i \gamma_{\mu} \partial^{\mu} - m_0 ) \psi 
  -{1\over 2} g_0 j^{\mu} j_{\mu}   \eqno{ (2.1)} $$
where the fermion current $  j_{\mu} $  is written as
$$  j_{\mu} = :\bar \psi  \gamma_{\mu} \psi :   . \eqno{ (2.2)}   $$
Choosing a basis where $\gamma_5$ is diagonal, the hamiltonian is written 
$$  H = \int dx \left[-i(\psi_1^{\dagger}{\partial\over{\partial x}}\psi_1
-\psi_2^{\dagger}{\partial\over{\partial x}}\psi_2 )+
m_0(\psi_1^{\dagger}\psi_2+\psi_2^{\dagger}\psi_1 )+
2g_0 \psi_1^{\dagger}\psi_2^{\dagger}\psi_2\psi_1 \right]  .  \eqno{(2.3)} $$

The hamiltonian eq.(2.3) can be diagonalized by the Bethe ansatz wave 
functions.
The Bethe ansatz wave functions satisfy the eigenvalue equation. 
However, they still do not have proper boundary conditions. The 
simplest way to define field theoretical models is to put the theory in a 
box of length $L$ and impose periodic boundary conditions (PBC) on the states. 

Therefore, we demand that $\Psi (x_1,..,x_N)$ be periodic in each argument 
$x_i$. This gives the boundary condition 
$$ \Psi (x_i=0) =  \Psi (x_i = L)   . \eqno{(2.4)} $$ 
This leads to the following PBC equations, 
$$ \exp (im_0 L \sinh \beta_i )= \exp (-i \sum_j \phi (\beta_i -\beta_j) ) 
   .   \eqno{(2.5)} $$ 
Taking the logarithm of eq.(2.5), we obtain 
$$ m_0 L \sinh \beta_i = {2\pi n_i} - \sum_j \phi (\beta_i -\beta_j)  
      \eqno{(2.6)} $$ 
where $n_i$'s are integer. These are equations which we should now solve.

First, we want to make a vacuum.  We write the PBC equations 
for the vacuum which is filled with negative energy particles 
( $\beta_i=i\pi -\alpha_i$ ),  
$$  \sinh \alpha_i = {2\pi n_i \over{L_0}}  
 - {2\over{L_0}} \sum_{j \not= i} \tan^{-1}\left[{1\over 2}g_0 
\tanh{1\over 2} (\alpha_i -\alpha_j) \right] ,  \qquad  
(i=1,..,N)   . \eqno{(2.7)} $$ 
 Now, $n_i$ runs as 
$$ n_i = 0, \pm 1, \pm 2, ..., \pm N_0    .  $$ 
We fix the values of $L_0$ and $N$, and then can solve eq.(2.7). 
This determines the vacuum. In this case, the vacuum energy $E_v$ can be 
written as  
$$ E_v =- \sum_{i=-N_0}^{N_0} m_0 \cosh \alpha_i  . \eqno{(2.8)} $$ 

Next, we want to make one particle-one hole $(1p-1h)$ 
 state. That is, we take out one 
negative energy particle ($i_0$-th particle) 
 and put it into a positive energy state. 
In this case, the PBC equations become 
\renewcommand{\theequation}{2.9\alph{equation}}
\setcounter{equation}{0}
\begin{eqnarray}
i\neq i_0 \nonumber \\
 & \sinh\alpha_i & = \frac{2\pi n_i}{L_0}-\frac{2}{L_0}\tan^{-1}
 \left[ {1\over 2}g_0\coth\frac{1}{2}(\alpha_i+\beta_{i_0})\right] 
 \nonumber \\
 & &- \frac{2}{L_0}\sum_{j\neq i,i_0}\tan^{-1}
\left[{1\over 2}g_0\tanh\frac{1}{2}
 (\alpha_i-\alpha_j)\right] \\
 \nonumber \\
i=i_0 \nonumber \\
 & \sinh\beta_{i_0} & = \frac{2\pi n_{i_0}}{L_0}+\frac{2}{L_0}\sum_{j\neq i_0}
 \tan^{-1}\left[{1\over 2}g_0\coth\frac{1}{2}(\beta_{i_0}+\alpha_j)\right]   
\end{eqnarray} 
where $\beta_{i_0}$ can be a complex variable as long as it can satisfy 
eqs.(2.9). 

These  PBC equations determine the energy of the one particle-one 
hole states which we denote by $E_{1p1h}^{(i_0)}$,  
$$ E_{1p1h}^{(i_0)}= m_0 \cosh \beta_{i_0} -
\sum_{\stackrel{\scriptstyle i=-N_0}{i\not= i_0}}^{N_0} 
m_0 \cosh \alpha_i  .  \eqno{(2.10)}   $$  

In the same way, we can set up the PBC equations for 2p$-$2h states. 

\vspace{2cm}

\item{\large Numerical results}

These PBC equations are solved numerically and we obtain the energies  
for the vacuum as well as the 1p$-$1h and the 2p$-$2h states. The bare 
numbers of the calculated energies are shown in Table 1. 

We now define the excitation energies with respect to the vacuum 
energy 
$$  \Delta E^{(0)}=E^{(0)}_{1p1h} -E_v $$ 
$$  \Delta E^{(1)}=E^{(1)}_{1p1h} -E_v   . $$ 
It turns out that these energies can be parametrized as 
 $$ \Delta E_{1p1h}^{(0)} = m_0 \left( A_0 + B_0 
 \left( {\rho\over m_0} \right)^{\alpha} \right)   \eqno{(3.1a)}  $$  
 $$ \Delta E_{1p1h}^{(1)} = m_0 \left( A_1 + B_1 
 \left( {\rho\over m_0} \right)^{\alpha} \right) .  \eqno{(3.1b)}  $$  
Now, we want to let $\rho \rightarrow \infty$, keeping 
$ \Delta E_{1p1h}^{(0)}$ and $ \Delta E_{1p1h}^{(1)}$ 
 finite. Since $\alpha$ is smaller than unity, 
we can make a fine-tuning of $m_0$ such that  
$$  m_0^{1-\alpha} \rho^{\alpha} = {\rm finite}  . $$  
In this case, we can identify the mass of the bound state $\cal M$ as  
$$ {\cal M} = 2m   \lim_{\rho \rightarrow \infty} 
\left( { \Delta E_{1p1h}^{(0)}\over{\Delta E_{1p1h}^{(1)} }} \right) = 
2m {B_0\over{B_1}}   .  \eqno{(3.2)}   $$  
Here $m$ is the physical fermion mass. 

In Table 2, we show our calculated result of the boson state 
as the function of the coupling constant $g$.  As can be seen from 
the Table 2, the calculated boson masses agree with those of 
the infinite momentum frame calculation by Fujita and Ogura. 

We have also calculated the 2p$-$2h configuration and found that 
there is no bound state in this configuration. Instead, the boson 
boson states appear as the continuum spectrum.  

\vspace{2cm}

\item{\large Finite size correction}

Since we solve the field theory in the box with its length $L$, 
we can calculate the finite size correction to the vacuum energy. 

For the massless case, the finite size correction is written as 
$$ \Delta E = -{\pi\over{6}} {c\over{L}} +...  $$
where $c$ denotes central charge and should be equal to 
unity for the Thirring model. In our calculations, 
the values of $c$ are found to be unity for the negative 
values of the coupling constant $g$. This is a good evidence 
that we solved the PBC equations properly.

\vspace{2cm}
\item{\large Discussions and further studies }

We have presented a new interpretation of the Bethe ansatz solutions of 
the massive Thirring model. Here, we solve the PBC equations 
directly but numerically without referring to the density of states 
or $string$ hypothesis. It is found that the Bethe ansatz solutions 
produce one bound state (a boson). This spectrum as the function of 
the coupling constant is consistent with Fujita-Ogura's solution. 

 Also, it is shown that the $string$ 
configurations taken by Bergknoff and Thacker do not satisfy the 
PBC equations and thus their $string$ is not a solution of the PBC 
equations.  In this way, the present result rules out a belief 
that the semiclassical 
result for the massive Thirring model is exact.

Also, the strong coupling expansion is performed in ref.[5] and 
the analytic expressions are obtained for the vaccum state energy 
as well as the boson boson scattering states. 
There, it turns out that the boson boson scattering states 
which are made of continuum states coincide with the twice of 
the boson mass. Therefore, we also learn 
from the strong coupling expansion that the $2p-2h$ states 
do not give any bound states.     

Now, we want to discuss the S-matrix method by Zamolodchikov and Zamolodchikov 
[6]. This factorized S-matrix method is also known to give the same spectrum 
as the semiclassical result for the sine-Gordon field theory or the massive 
Thirring model. 
 Concerning the factorization of the S-matrix 
 for the particle-particle scattering 
in the massive Thirring model, 
one may  convince oneself that the
factorization is indeed satisfied. 

However, there is a serious problem for the S-matrix factorization 
of the particle hole scattering. The problem is that the rapidity 
variables determined for $n-$particle $n-$hole states are different 
from each other as well as 
 those determined for the vacuum. Since the Lagrangian of the 
massive Thirring model satisfies the charge conjugation, 
one tends to believe that the crossing symmetry should be automatically 
satisfied. Indeed, the crossing symmetry itself should hold. 
However, we should be careful whether the crossing symmetry can 
commute with the factorization of the S-matrix or not. 
Recent calculations in ref.[7] show that the crossing symmetry 
and the factorization of the S-matrix do not commute with eath other. 
Therefore, it turns out that the S-matrix factorization 
for the particle hole scattering does not hold. In a sense, it is 
reasonable that the S-matrix factorization is consistent with the 
semiclassical results since it is indeed due to the consequence 
of the neglect of the operartor commutability.

\newpage
\begin{center}
\underline{Table 1} \\
\ \ \\
 $N$=1601 \ \ \  $L_0$=100 \ \ \ \ \ \  \\
\ \ \\ 
\begin{tabular}{|l|c|c|c|c|}
\hline
 & \multicolumn{1}{|l}{$\begin{array}{lcl}
   \frac{g}{\pi}& = &1 \\
   \end{array}$} 
 & \multicolumn{1}{|l}{$\begin{array}{lcl}
   \frac{g}{\pi} & = & 1.25 \\
   \end{array}$} 
 & \multicolumn{1}{|l}{$\begin{array}{lcl}
   \frac{g}{\pi} & = & 1.5 \\
   \end{array}$} 
 & \multicolumn{1}{|l|}{$\begin{array}{lcl}
   \frac{g}{\pi} & = & 1.7 \\
   \end{array}$} \\   
\hline
&&&& \\
$E_v$ & $- 9095.31$ & $-6215.70$ & $-4205.83$ & $-2995.13$ \\
\hline
&&&& \\
$E^{(0)}_{1p1h}$ & $- 9089.43$ & $-6210.69$ & $-4201.76$ & $-2991.83$ \\
%&&&& \\
\hline
&&&& \\
$E^{(1)}_{1p1h}$ & $- 9080.78$ & $-6197.08$ & $-4182.54$ & $-2966.95$ \\
&&&& \\
$E^{(2)}_{1p1h}$ & $- 9080.72$ & $-6197.02$ & $-4182.48$ & $-2966.89$ \\
&&&& \\
$E^{(3)}_{1p1h}$ & $- 9080.66$ & $-6196.96$ & $-4182.42$ & $-2966.82$ \\
&&&& \\
$E^{(4)}_{1p1h}$ & $- 9080.59$ & $-6196.90$ & $-4182.35$ & $-2966.76$ \\
&&&& \\
$E^{(5)}_{1p1h}$ & $- 9080.53$ & $-6196.83$ & $-4182.29$ & $-2966.70$ \\
&&&& \\
$E^{(6)}_{1p1h}$ & $- 9080.47$ & $-6196.77$ & $-4182.23$ & $-2966.64$ \\
\hline
\hline
&&&& \\
$E^{(1,-1)}_{2p2h}$ & $- 9066.23$   & $-6178.55$ & $-4159.13$ & $-2938.79$  \\
&&&& \\
$E^{(2,-2)}_{2p2h}$ & $- 9066.10$   & $-6178.43$ & $-4158.99$ & $-2938.66$  \\
&&&& \\
$E^{(3,-3)}_{2p2h}$ & $- 9065.97$   & $-6178.30$ & $-4158.86$ & $-2938.52$  \\
&&&& \\
$E^{(4,-4)}_{2p2h}$ & $- 9065.85$   & $-6178.17$ & $-4158.74$ & $-2938.38$  \\
&&&& \\
$E^{(5,-5)}_{2p2h}$ & $- 9065.72$   & $-6178.04$ & $-4158.61$ & $-2938.25$  \\
&&&& \\
$E^{(6,-6)}_{2p2h}$ & $- 9065.59$   & $-6177.91$ & $-4158.49$ & $-2938.12$  \\
\hline
\end{tabular}
\\
\vspace*{1cm}  

\begin{minipage}{13cm}
We plot the calculated energies of $E_v$, $E_{1p1h}^{(n)} \  (n=0,6)$ 
and $E_{2p2h}^{(n,-n)} \  (n=1,6)$ for some values of the coupling 
constant $g\over{\pi}$ 
with the fixed $L_0=100$. The number of particles here 
is $N=1601$. Note that we put $m_0 =1$ in our calculations.  
\end{minipage}
\end{center}

\newpage
\begin{center}
\underline{Table 2}  \hspace{1.5cm}  \\
\ \\
\begin{tabular}{|c|c|c|c|c|c|}
\hline
 & & \multicolumn{1}{l|}{Present} & \multicolumn{1}{l|}{Fujita}  
 & \multicolumn{1}{l|}{Dashen et al.} & \multicolumn{1}{l|}{Bergknoff} \\
 & & \multicolumn{1}{l|}{Calculation} & \multicolumn{1}{l|}{Ogura} & 
 & \multicolumn{1}{l|}{Thacker} \\
\hline
&&&&& \\
${\cal M}$ & $\displaystyle{\frac{g}{\pi}=0.8}$ 
& $1.01  m$ & $0.98m$ & $0.83m$ & $0.51m$ \\
& ($g_0=4.19$) &&&& \\
\hline
&&&&& \\
${\cal M}$ & $\displaystyle{\frac{g}{\pi}=1.0}$ 
& $0.75  m$ & $0.77m$ & $0.62m$ & $0.34m$ \\
& ($g_0=6.28$) &&&& \\
\hline 
&&&&& \\
${\cal M}$ & $\displaystyle{\frac{g}{\pi}=1.25}$ 
& $0.47  m$ & $0.54m$ & $0.41m$ & $0.20m$ \\
& ($g_0=10.5$) &&&& \\
\hline 
\end{tabular}

\vspace{2cm}
\begin{minipage}{13cm}
We plot the predicted values of the boson mass $\cal M$ by the present 
calculation, by the infinite momentum frame calculation ( Fujita-Ogura ), 
by the semiclassical method ( Dashen et al.) and by the Bethe ansatz 
technique  with $string$ hypothesis ( Bergknoff - Thacker ). 
\end{minipage}
\end{center}

\vspace{1cm}
{\large Reference}

%{\large REFERENCES }
\baselineskip = 8 mm

1.  R. F. Dashen, B. Hasslacher and A. Neveu, 
Phys. Rev. {\bf D11} (1975), 3432 

2. H. Bergknoff and H.B. Thacker, Phys. Rev. Lett. {\bf 42} (1979), 135  

3. T. Fujita and A. Ogura,  Prog. Theor. Phys. {\bf 89} (1993), 23 

4. W. Thirring, Ann. Phys. (N.Y) {\bf 3} (1958), 91 

5. T. Fujita, C. Itoi and H. Mukaida, to be published. 

6. A.B.Zamolodchikov and A.B.Zamolodchikov, Ann. Phys. {\bf 120} (1979), 253 

7. T. Fujita and M. Hiramoto, to be published.

\end{enumerate}
\end{document}